# Size-dependent yield hardness induced by surface energy

**Yue Ding** and **Gang-Feng Wang***

Department of Engineering Mechanics, SVL, Xi'an Jiaotong University, Xi'an 710049, China

E-mail: wanggf@mail.xjtu.edu.cn

**Abstract**

Size dependent hardness has long been reported in nanosized indentations, however the corresponding explanation is still in exploration. In this paper, we examine the influence of surface energy on the hardness of materials under spherical indentation. To evaluate the ability of materials to resist indentation, a yield hardness is defined here as the contact pressure at the inception of material yield. It is found that this defined hardness is an intrinsic material property depending only on the yield strength and Poisson's ratio in conventional continuum mechanics. Then, the impact of surface energy on the yield hardness is analyzed through finite element simulations. By using the dimensional analysis, the dependences of the yield hardness and critical indent depth at yield initiation on surface energy have been achieved. When the yield strength is comparable to the ratio of surface energy density to indenter radius, surface energy will alter the yield hardness and the critical indent depth. As the size of indenter decreases to nanoscale, both the yield hardness and the indent depth will increase significantly. This study provides a possible clarification to the size dependence of hardness and a potential approach to measure the yield strength and surface energy of solids through nanosized indentations.

**Key words**: hardness, nanoindentation, surface energy

# 1. Introduction

With the development of depth-sensing indentation technique, nanoindentation tests are widely conducted to extract the mechanical properties of materials, such as elastic modulus and indentation hardness [1-4]. In this technique, the hardness is usually defined as the indentation load divided by the projected contact area [4]. However, this value will vary in the indentation process, either a specified load or an indent depth has to be chosen as a standard. As another alternative, if the maximum contact pressure within the whole indentation process is employed to define the hardness, it depends only on the mechanical properties of measured materials. For the spherical indentation of elastic-ideally plastic materials, the hardness is related to the yield stress and elastic modulus, and is approximately three times of yield stress for materials with large ratio of elastic modulus to yield stress [5-7]. Thus the indentation hardness is not an independent material property but an indicator or combination of other mechanical properties.

In macroscopic continuum mechanics, the mechanical properties of solids are expected to be independent of sample size. However, when the indent depth decreases to micrometers or even nanometers, the measured hardness displayed clear size dependence [8-10]. Various factors have been ascribed to the causes of size dependent hardness [11]. For micro-indentations, strain gradient theories are successful to interpret the size effects of measured hardness [12, 13], but not sufficient for nano-indentatons [14]. Atomistic simulations [15] and experimental evaluations [9] strongly suggested that surface effect plays a critical role in nanosized contact problems [16].

In recent decades, attentions on the influence of surface tension or surface energy on contact problems of elastic materials has grown considerably. Based on the surface Green's functions with surface tension [17, 18], Long et al. [19, 20] analysed the plane strain and axisymmetric contact problems respectively, and suggested that the existence of surface tension could lead to evident size-dependent contact responses. Using Stroh's formalism and considering the effects of surface stress and surface elasticity, Hayashi et al. [21] investigated the contact problems of anisotropic elastic half space, the results of load-depth relation agreed well with that of molecular dynamics analysis. Through FEM simulations, Ding et al. [22] studied the nanoindentation of soft solids with surface energy. The obtained overall indentation responses, i.e. the load-depth relations, were then employed to explain the depth dependent moduli of biological cells [23]. The contact problems with both surface tension and adhesion have been considered by Hui and his coworkers [24, 25]. However, the effect of surface energy on material hardness has been rarely addressed.

The hardness in most previous studies focused on plastic deformation at large indentation, in fact, the contact pressure at the inception of yield could also be an indicator to characterize the mechanical property of materials. In the present work, a yield hardness is defined as the mean contact pressure at the onset of yield, which is found to be dependent on the yield stress and Poisson's ratio and thus can also be used to indicate the resistant ability to indentation. Then we consider the effects of surface energy on the yield hardness under spherical indentation. Through finite element simulations and dimensional analysis, the expressions of the yield hardness and the indent depth at yield initiation are generalized explicitly.

**2. Finite element model**

As illustrated in Fig. 1, an isotropic half space is indented by a rigid sphere with radius $R$. An external load $P$ is applied on the indenter, and induces a circular contact area with radius $a$ and an indent depth $d$. Refer to a cylindrical coordinate system, with the origin located at the initial contact point and the $z$-axis directed to the compressive direction.

Our FEM simulations are carried out in the commercial software ABAQUS. Through the user subroutine UEL, we develop surface elements to incorporate the influence of surface energy. In our simulations, a constant surface energy density $\gamma$ is assumed on the surface. To seek the equilibrium state of this contact problem, the Newton-Raphson method is adopted in ABAQUS. Further details about finite element simulations incorporating surface energy can be referred to reference [26].

An axisymmetric model is adopted here to simulate the spherical indentation on a half space, as shown in Fig. 2. The isotropic half space is discretized by four-node linear axisymmetric quadrilateral elements (CAX4), and surface energy is accounted by user-defined surface elements. The displacement along the $z$-direction is prescribed to be zero on the bottom boundary, and the displacement along the $r$-direction is fixed on the symmetrical axis ($r=0$). The contact condition is set as frictionless between the half space and the indenter. As preparation, convergence has been examined to guarantee the accuracy of our numerical results.

## 3. Results and discussions

For the indentation of an elastic half space by a rigid spherical indenter, the classic contact theory predicts the stresses along the $z$-axis as [27]

$$\sigma_r = \sigma_\theta = \frac{3}{2} p_m \left\{ -(1+v)\left[1 - \frac{z}{a}\arctan\left(\frac{a}{z}\right)\right] + \frac{1}{2}\left(1 + \frac{z^2}{a^2}\right)^{-1} \right\}, \tag{1}$$

$$\sigma_z = \frac{3}{2} p_m \left\{ -\left(1 + \frac{z^2}{a^2}\right)^{-1} \right\} \tag{2}$$

$$\tau_{rz} = \tau_{z\theta} = \tau_{\theta r} = 0 \tag{3}$$

where $p_m = P/(\pi a^2)$ is the mean pressure over the contact region and $\sigma_r$, $\sigma_\theta$ and $\sigma_z$ are the principal stresses. Accordingly, the equivalent von Mises stress $\sigma_v$ can be expressed as

$$\sigma_v = \sqrt{(\sigma_r - \sigma_z)^2} = \frac{3}{2} p_m \left\{ \frac{3}{2}\left(1 + \frac{z^2}{a^2}\right)^{-1} - (1+v)\left[1 - \frac{z}{a}\arctan\left(\frac{a}{z}\right)\right] \right\} \tag{4}$$

For this contact problem, classical contact mechanics demonstrates that the Mises stress inside the indented material achieves its maximum value on the $z$-axis at certain distance beneath the surface [27]. The maximum Mises stress $\sigma_v^{max}$ and the location of yield initiation $z_0$ can be obtained by seeking the maximum value of Eq. (4). Through numerical calculation, the dependences of normalized maximum Mises stress $\sigma_v^{max}/p_m$ and normalized position $z_0/a$ on Poisson's ratio are plotted in Fig. 3, which can be well fitted by the following expressions

$$\frac{\sigma_v^{max}}{p_m}=1.134-0.674\nu \tag{5}$$

$$\frac{z_0}{a}=0.381+0.334\nu \tag{6}$$

It shows that as the Poisson's ratio increases, the value of $\sigma_v^{max}/p_m$ decreases and the relative position $z_0/a$ becomes deeper. These relations hold for elastic deformation before yield.

In this work, a yield hardness $H_y$ is defined as the contact pressure $p_m$ when the maximum Mises stress in the indented solid reaches the yield strength $Y$ of material. Thus, for the case without surface energy, the yield hardness $H_{y0}$ can be achieved by

$$H_{y0}=\frac{Y}{1.134-0.674\nu} \tag{7}$$

This implies that the yield hardness defined here is an intrinsic property, which only relates to the yield strength and Poisson's ratio of the material, and thus can be used to indicate its resistance ability to deformation.

Using the indentation load-depth relationship of the Hertzian theory, that is

$$P=\frac{4}{3}E^*R^{0.5}d^{1.5} \tag{8}$$

the critical indent depth $d_0$ at the onset of yield can be derived as

$$d_0=R\left[\frac{\pi}{(1.512-0.899\nu)}\frac{Y}{E^*}\right]^2 \tag{9}$$

where $E^*=E/(1-\nu^2)$ is the combined elastic modulus.

Now we examine influence of surface energy on the stress distributions of solids under spherical indentation. In our simulations, we consider Ag material with elastic modulus $E$=83GPa, Poisson's ratio $v$=0.37 and $\gamma$=7.2J/m$^2$ [28]. For indentations under a fixed load $P$=10nN by various sizes of spherical indenters ($R$=20 nm, 100 nm and 1 μm), Figs. 4 and 5 display the distribution of normalized stresses $\sigma_r$ and $\sigma_z$ along the $z$-axis, respectively. For the cases without surface energy, our FEM simulation results (the triangular dots) coincide with the classical solution given by Eqs. (1) and (2). It can be found that the existence of surface energy decreases both stresses $\sigma_r$ and $\sigma_z$ along the $z$-axis. With the decrease of the indenter radius, the difference of stresses between Hertzian solution and the cases with surface energy becomes evident, which indicates a significant surface effect.

Figure 6 demonstrates the distribution of normalized von Mises stress $\sigma_v$ along $z$-axis. For Ag material with $v$=0.37, the classical contact mechanics predicts that the maximum Mises stress ($\sigma_v/p_m$=0.8826) occurs at $z/a$=0.5042. It is seen that our FEM simulations without surface energy coincides with this theoretical prediction, as shown in Fig. 6. When surface energy is considered, the Mises stress decrease along the $z$-axis. As the indenter size becomes smaller, the stress distribution deviates further from the classical solution and its peak value decreases more remarkably.

Then we analyze the influence of surface energy on the yield hardness and the location of the yield inception. According to dimensional analysis, the influence of surface energy on the yield hardness $H_y$ can be expressed by

$$H_y = H_{y0} f\left(\frac{YR}{\gamma}\right) \tag{10}$$

where $f$ is a function of the dimensionless parameter $YR/\gamma$.

For various indenter radii and elastic materials with different elastic moduli, Poisson's ratios and surface energy densities, finite element simulations are performed to achieve the relation between yield hardness and yield strength. Fig. 7 displays the dependence of normalized yield hardness on the ratio $YR/\gamma$, and the relation can be well reproduced by the function

$$\frac{H_y}{Y} = \frac{1}{1.134 - 0.674\nu}\left[1 + 1.92\left(\frac{YR}{\gamma}\right)^{-0.863}\right] \tag{11}$$

The explicit expression indicates that when yield strength $Y$ is comparable to or smaller than the ratio $\gamma/R$, surface energy would affect the yield hardness. As the size of indenter decreases, corresponding to the decrease of the ratio of $YR/\gamma$, the yield hardness would increase compared to its macroscopic counterpart. For cases with the ratio $\gamma/R$ much larger than $Y$, the surface energy effect is negligible, and the yield hardness approaches a constant value given by Eq. (7).

The influence of surface energy on the critical indent depth at yield initiation has also been investigated. For different indenter sizes and various materials, the FEM results of normalized indent depth with respect to the ratio $YR/\gamma$ are plotted in Fig. 8. The relation between critical indent depth $d_s$ and ratio $YR/\gamma$ can be formulated as

$$\frac{d_s}{d_0} = \left[1 + 0.213\left(\frac{YR}{\gamma}\right)^{-0.59}\right] \tag{12}$$

When yield strength $Y$ is much larger than the ratio $\gamma/R$, the surface effect vanishes and the critical indent depth $d_s$ can be directly obtained by Eq. (9). For the smaller ratios of $YR/\gamma$, thus the more significant surface effect, the normalized indent depth

$d_s/d_0$ increases, indicating that the a larger indent depth is required to reach the yield due to surface energy. This implies that surface energy tends to resist yield. The above explicit expressions (Eqs. (11) and (12)) can be used to determine the yield strength and surface energy of solids through nanoindentations.

## 4. Conclusions

The spherical indentation on an elastic half space with surface energy has been investigated. The yield hardness, defined as the contact pressure when the yield initiates, is proved to be an intrinsic mechanical property of material. Based on finite element simulations and dimensional analysis, the influences of surface energy on the yield hardness and the critical indent depth of initial yield have been examined. When the yield strength is comparable to or smaller than the ratio of surface energy density to the indenter radius, both the yield hardness and the critical indent depth would increase comparing with cases without surface energy. The present results are beneficial for understanding the size effect of hardness and provide a more accurate avenue to quantify the yield strength and surface energy of materials by nanosized indentation.

**Acknowledgements**

Supports from the National Natural Science Foundation of China (Grant No. 11525209) and China Postdoctoral Science Foundation are acknowledged.

**Figure captions:**

Fig. 1. Spherical indentation on an isotropic half space.

Fig. 2. Finite element models of the spherical indentation.

Fig. 3. The dependence of normalized maximum Mises stress and normalized location of initial yield point on Poisson's ratio.

Fig. 4. Distribution of normalized principal stresses $\sigma_r$ along the $z$-axis.

Fig. 5. Distribution of normalized principal stresses $\sigma_z$ along the $z$-axis.

Fig. 6. Distribution of normalized Mises stress $\sigma_v$ along the $z$-axis.

Fig. 7. Relation between yield hardness and yield strength for elastic half space with surface energy.

Fig. 8. Relation between critical indent depth and yield strength for elastic half space with surface energy.

**Figures:**

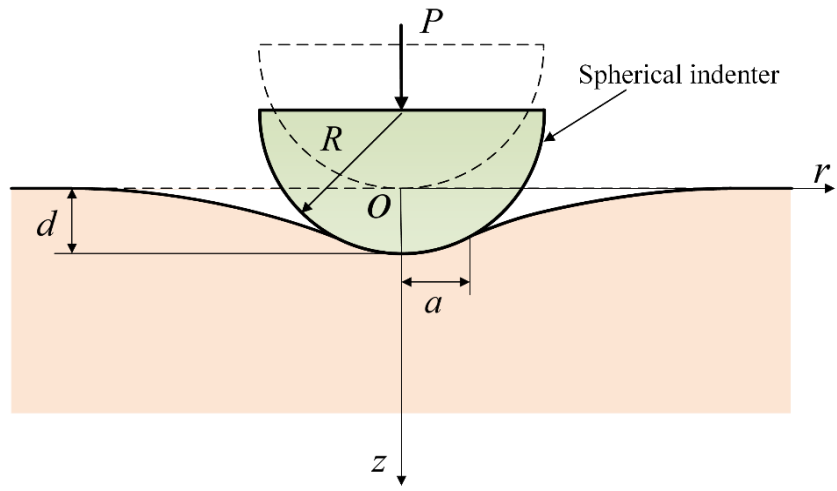

Fig. 1

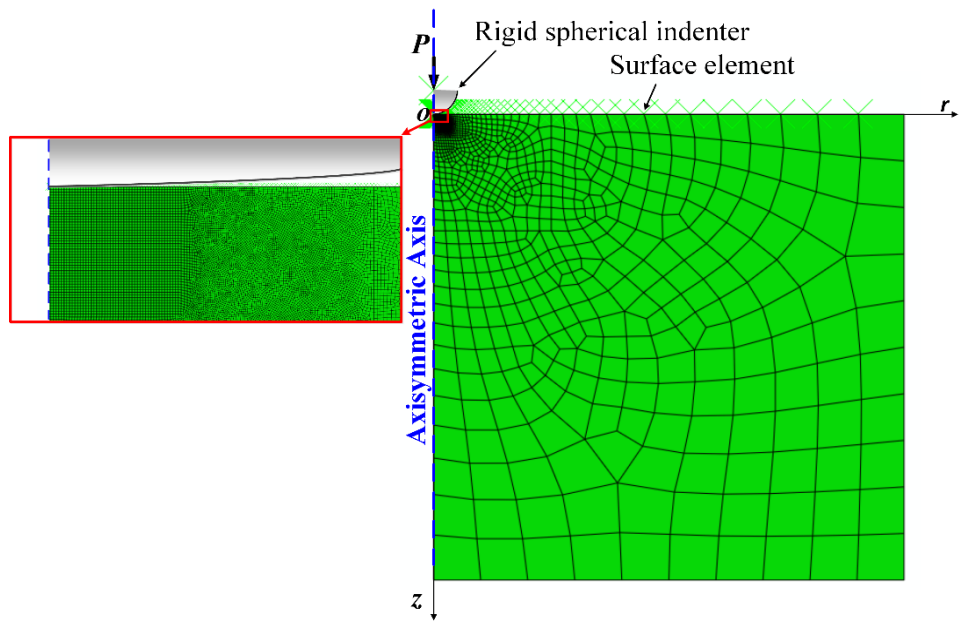

Fig. 2

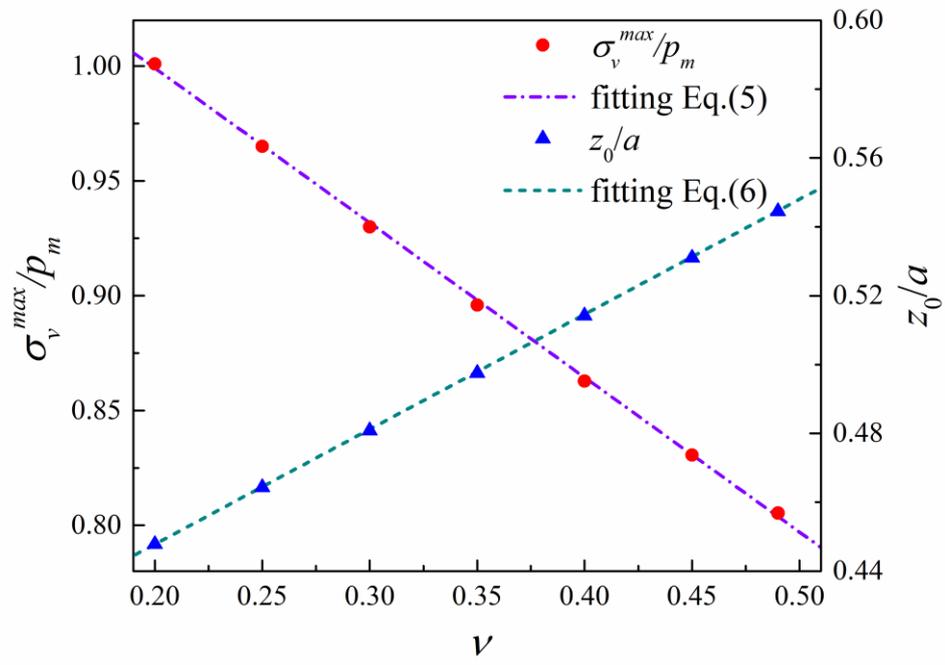

Fig. 3

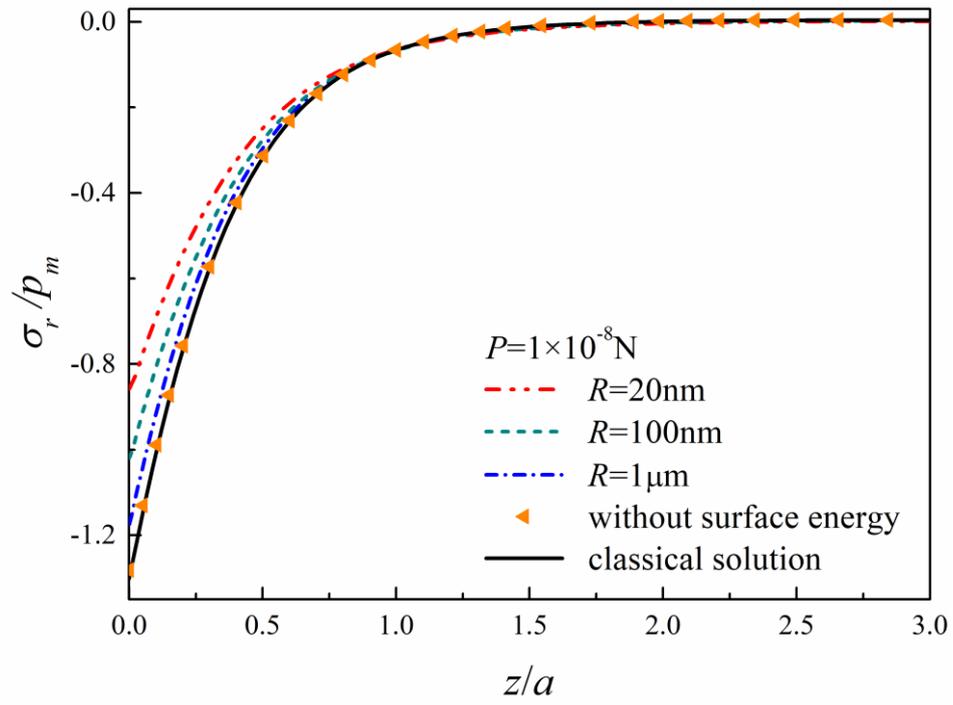

Fig. 4

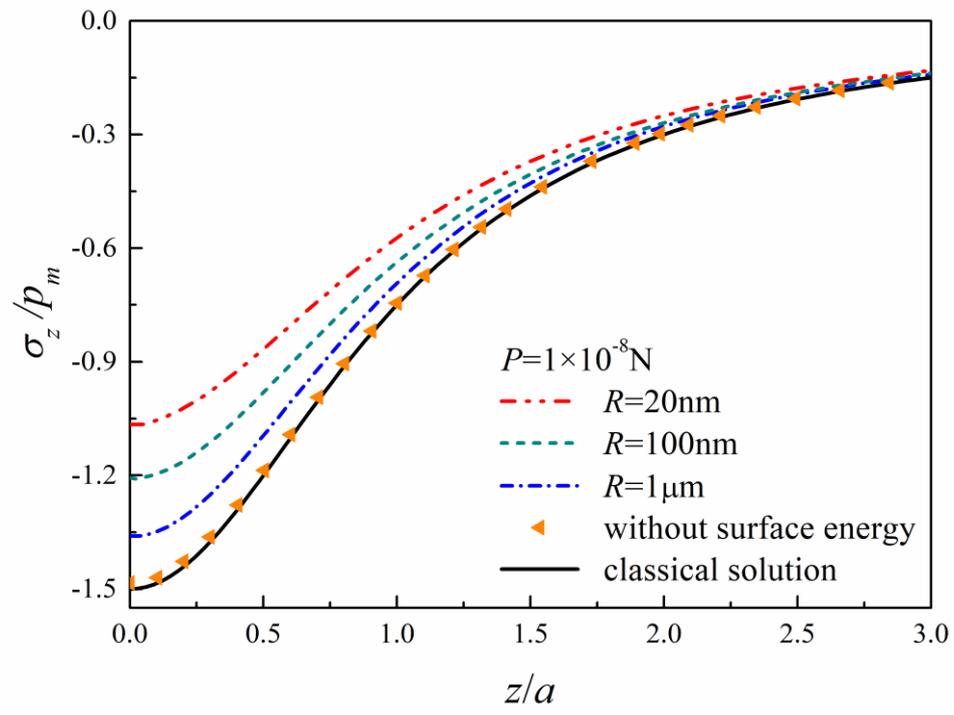

Fig. 5

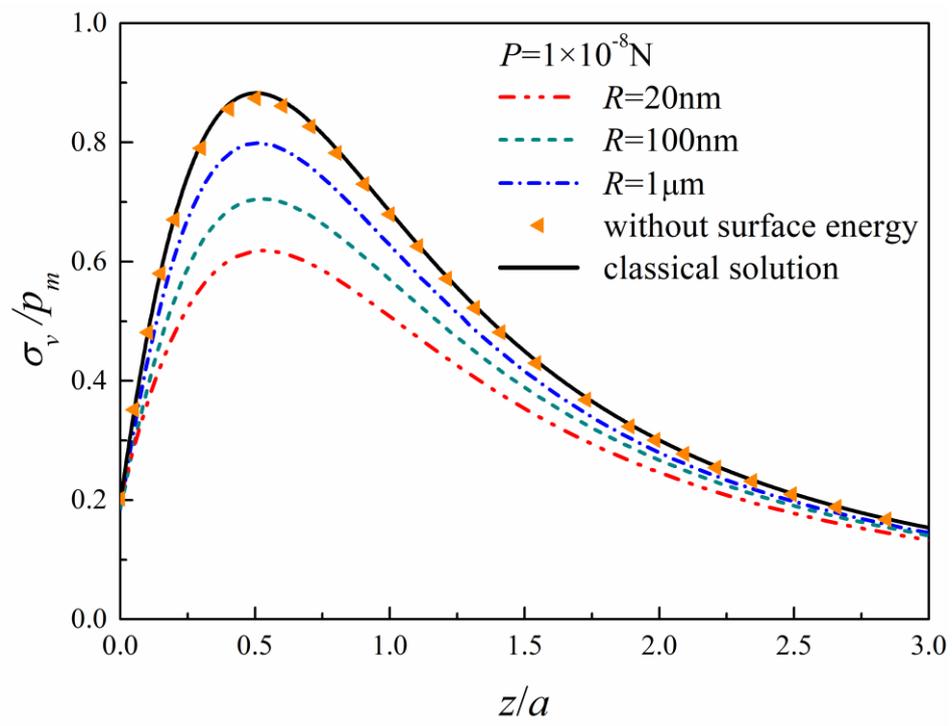

Fig. 6

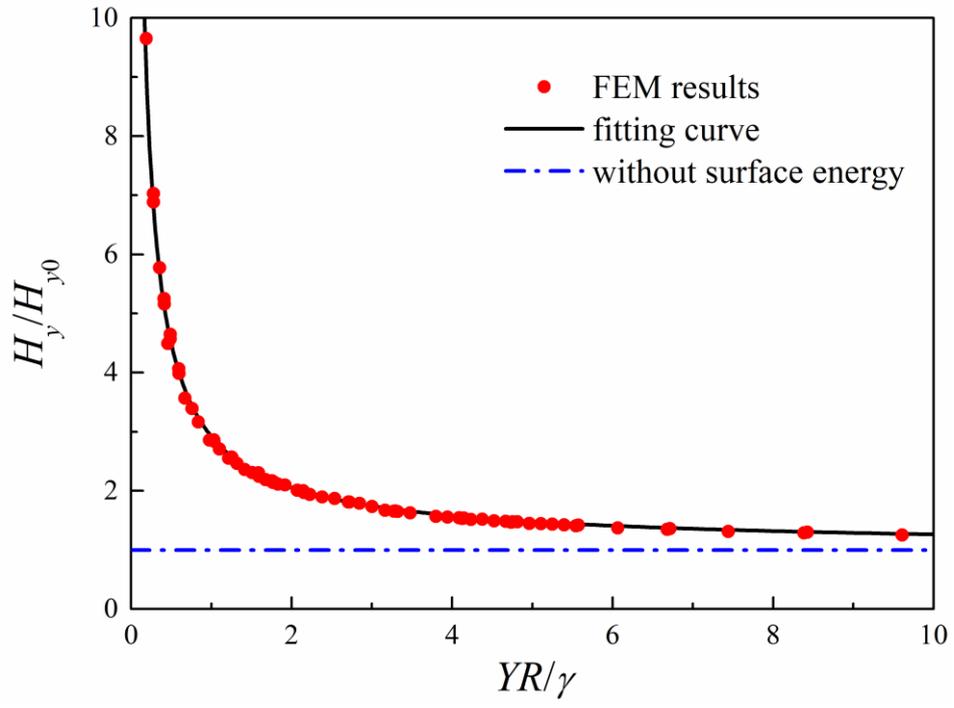

Fig. 7

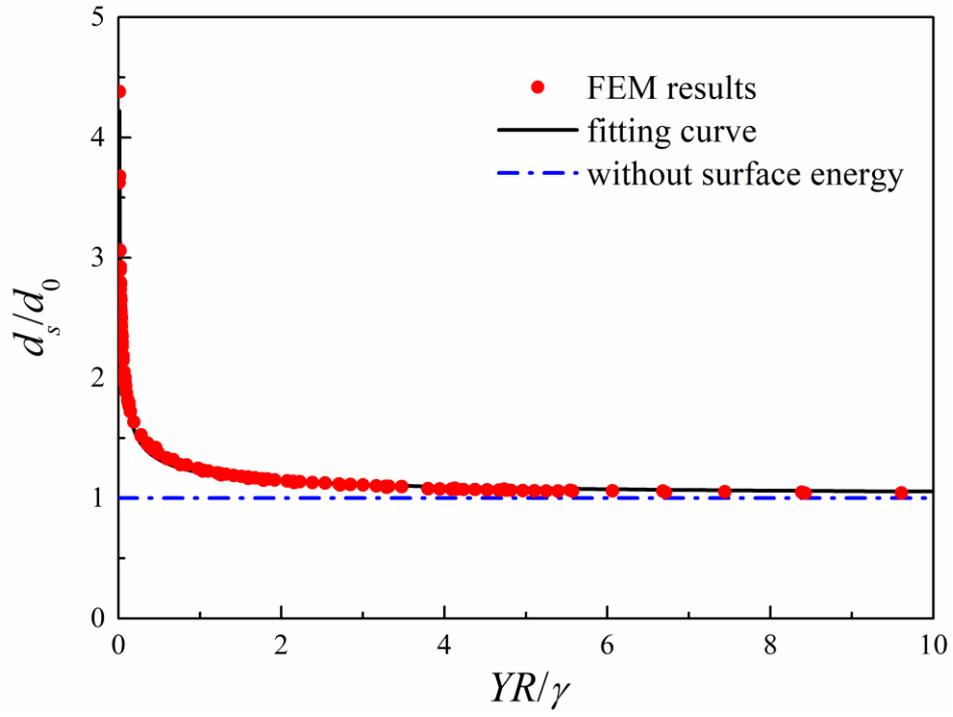

Fig. 8